\documentclass[twocolumn,pre,floatfix]{revtex4}
\usepackage{psfrag,epsfig,amsfonts,amssymb,amsmath,wasysym,bm}
\usepackage{dcolumn}
\usepackage{bbm,psfrag}

\usepackage[normalem]{ulem}
\usepackage{color}

\newcommand{\mic}{\mathrm{mc}}
\newcommand{\mini}{\mathrm{min}}
\newcommand{\maxi}{\mathrm{max}}
\newcommand{\QQ}{P}
\newcommand{\amic}{a_{\mathrm{mc}}}
\newcommand{\id}{\ensuremath{\mathbbm{1}}}
\newcommand{\rhomic}{\rho_{\mathrm{mc}}}

\newcommand{\HA}{\mbox{HA}}
\newcommand{\HV}{\mbox{HV}}
\newcommand{\hr}{{\cal H}}
\newcommand{\tr}{\mbox{Tr}}
\newcommand{\da}{\Delta_{\! A}}

\providecommand{\norm}[1]{\|#1\|}

\newcommand{\RR}{{\mathbb R}}
\newcommand{\CC}{{\mathbb C}}

\newcommand{\pr}{{\rm{Prob}}}

\begin{document}

\title{Dynamical typicality of isolated many-body 
quantum systems}

\author{Peter Reimann}
\affiliation{Fakult\"at f\"ur Physik, 
Universit\"at Bielefeld, 
33615 Bielefeld, Germany}

\begin{abstract}
Dynamical typicality refers to the property that
two pure states, which initially exhibit (almost) 
the same expectation value 
for some given 
observable $A$, are very likely to exhibit also 
very similar expectation values when evolving in time
according to the pertinent Schr\"odinger 
equation.
We unify and generalize a variety of 
previous findings of this type
for sufficiently high dimensional quantum 
mechanical model systems.
Particular emphasis is put on the necessary 
and sufficient conditions, which the 
initial expectation value 
and the spectrum of
$A$ have to fulfill.
\end{abstract}

\maketitle

\section{Introduction}
\label{s1}
The following quite remarkable feature
of isolated, high dimensional quantum 
systems has been discovered and named
``dynamical typicality'' in a hallmark 
paper by Bartsch and Gemmer \cite{bar09}:
The vast majority of all pure states 
with very similar expectation values 
of some observable at a given 
initial time, will yield very similar expectation 
values of the same observable also at any 
later time.
Most notably, this prediction does not 
depend on any details of the Hamiltonian
which governs the time evolution.
However, the 
restrictions in Ref. \cite{bar09}
regarding the initial expectation
value and also the considered 
observable itself were still quite
significant.
In an entirely unrelated work,
M\"uller, Gross, and Eisert
explored by means of concentration
of measure concepts the typical 
properties of random pure states 
with a fixed expectation value 
of some observable \cite{mul11}.
Somewhat similar investigations were 
independently carried out even earlier
by Fine in Ref. \cite{fin09}.
Yet another related finding
is contained in the last paragraph 
of Ref. \cite{rei07}:
If pure initial states are randomly sampled 
according to any statistical ensemble that 
corresponds to a density operator of low purity, 
then most of them exhibit very 
similar expectation values at 
any later time.
A less general version of the same result
was also established and numerically 
exemplified in Ref. \cite{alv08}.

The main objective of the present 
work is to unify and extend all those
previously independent approaches
and results.
In particular, Ref. \cite{mul11}
closes with ``the hope that methods similar to the 
ones established here help assessing questions 
of typicality in the context of quantum dynamics 
and addressing key open problems in the theory 
of relaxation of non-equilibrium complex quantum 
systems.''
Progress along these lines is exactly
at the focus of our present explorations.

\section{General framework}
\label{s2}
We consider an isolated quantum system,
which is described by means of some Hilbert 
space $\hr$ of large but finite dimension
$N$, and whose dynamics is governed by 
an arbitrary Hamiltonian $H:\hr\to\hr$.
Choosing any normalized $|\phi\rangle\in\hr$ 
as initial state (at $t=0$), 
its time evolution can be 
written as $|\phi(t)\rangle=U_t|\phi\rangle$, 
where $U_t:=e^{-iHt/\hbar}$.
Going over from the Schr\"odinger to 
the Heisenberg picture of quantum mechanics,
the resulting expectation value for an 
arbitrary observable $O$ at time $t$ 
thus follows as
\begin{eqnarray}
& & 
\langle \phi(t)|O|\phi(t)\rangle
=\langle\phi|O_t|\phi\rangle
\ ,
\label{1}
\\
& & O_t := U_t^\dagger O\, U_t \ .
\label{2}
\end{eqnarray}

The main objective of our paper is
to establish dynamical typicality 
in the following sense: 
For the vast majority of 
all initial states $|\phi\rangle\in \hr$,
which exhibit (almost) identical 
expectation values
$\langle\phi|A|\phi\rangle$
for some given observable $A$,
also the expectation values for  
any arbitrary but fixed observable 
$O$ and time point $t$ in (\ref{1})
will be very similar to each other.
Unlike in Ref. \cite{bar09}, the 
two observables $A$ and $O$ are
thus not any more required to be 
identical.
A more precise meaning of the term 
``vast majority'' will be provided later.
We also note that for any given $t$ 
and $O$, there may still be a 
small set of ``untypical'' initial states 
$|\phi\rangle$, entailing significantly
different expectation value 
in (\ref{1}).
Moreover, this set will usually
be different for different time 
points $t$ and/or different
observables $O$.

Quantitatively, the temporal evolution
of the expectation values in (\ref{1})
as well as the above mentioned
$t$- and $O$-dependent sets of ``untypical'' 
$|\phi\rangle$'s are determined in a
very complicated way by many details 
of the Hamiltonian $H$.
Given that $H$ can still be 
chosen arbitrarily (see above (\ref{1})), 
it is obvious that apart from 
establishing dynamical typicality 
{\em per se}, no further
conclusion regarding the actual time
evolution and the untypical sets
will be possible.

Note that {\em a priori}, the pertinent Hilbert 
space $\hr_0$ of a given model system is often 
of infinite dimension, and the actual Hamiltonian 
and observables are Hermitian operators on that
Hilbert space $\hr_0$.
Our above introduced, $N$ dimensional Hilbert 
space $\hr$ then represents, for instance, 
an energy shell, i.e., it is spanned by $N$ 
eigenvectors of the original Hamiltonian, 
whose eigenvalues are contained in an energy 
interval, which is microscopically large 
(thus $N$ is large) but macroscopically 
small (well defined system energy).
Since we are only interested in vectors
$|\phi\rangle$ with support in $\hr$, also 
the support of 
the Hamiltonian and the observables 
can be readily projected (restricted) from 
$\hr_0$ to $\hr$ without any change of 
the dynamics and expectation values.
More generally, $\hr$ may be any high
but finite dimensional subspace of 
$\hr_0$ which is invariant under 
the original system Hamiltonian.
The straightforward and well known
formal details of such a transition 
from $\hr_0$ to $\hr$ are provided,
e.g., in Ref. \cite{rei15}.

\section{Main idea and result}
\label{s3}
%
\subsection{Basic definitions and properties}
\label{s31}
The observable $A$ introduced below (\ref{2})
can be rewritten in terms of its
eigenvalues $a_n$ and eigenvectors 
$|n\rangle$ as
\begin{equation}
A=\sum_{n=1}^N a_n\, |n\rangle\langle n| 
\ .
\label{3}
\end{equation}
Furthermore, the largest and 
smallest eigenvalues of $A$ will be denoted 
as $a_{\maxi}$ and $a_{\mini}$, respectively.
Their difference
\begin{eqnarray}
\da := a_{\maxi}-a_{\mini}
\label{3a}
\end{eqnarray}
thus represents the measurement 
range of the observable $A$.
Finally, we define the microcanonical ensemble as
\begin{eqnarray}
\rhomic:=
\frac{1}{N}\sum_{n=1}^N|n\rangle\langle n|
\label{37}
\end{eqnarray}
and the concomitant microcanonical 
expectation value of $A$ from (\ref{3}) as
\begin{eqnarray}
a_{\mic} & := & \tr\{\rhomic A\}
= \frac{1}{N}\sum_{n=1}^N a_n \ .
\label{38}
\end{eqnarray}
From now on, we tacitly exclude the trivial case
$a_{\mini}=a_{\maxi}$.
It thus follows that
$a_{\mini}<\amic<a_{\maxi}$.

The main objective of the present 
Sec. \ref{s3} will be to construct an 
ensemble of normalized 
vectors $|\phi\rangle\in\hr$, 
most of which exhibit expectation values 
of $A$ very close to some preset value 
$a$, and which furthermore entail
dynamical typicality properties.
Our starting point in constructing this 
ensemble consists in the
following key observation:
Given an arbitrary but fixed 
real number
\begin{equation}
a\in(a_{\mini},a_{\maxi}) \ ,
\label{4}
\end{equation}
there exists a unique 
$y\in\RR$ so that the quantities
\begin{eqnarray}
p_n & := & \frac{1}{N}\frac{1}{1+y\,(a-a_n)}
\label{5}
\end{eqnarray}
have the properties that
\begin{eqnarray}
& & p_n > 0 \ \mbox{for all} \ n=1,...,N\, ,
\label{6}
\\
& & \sum_{n=1}^N p_n = 1\ ,
\label{7}
\\
& & \sum_{n=1}^N a_n p_n = a \ .
\label{8}
\end{eqnarray}

The derivation of these properties
is provided in full detail in 
Appendix A, and can be summarized 
as follows:
Considering the function
\begin{eqnarray}
g(x) & := & \frac{1}{N}\sum_{n=1}^N \frac{1}{1+x(a-a_n)}
\ ,
\label{8a}
\end{eqnarray}
one readily verifies that $g(x)\to\infty$ 
as $x$ approaches $x_{\maxi}:=1/(a_{\maxi}-a)>0$ 
from below, and likewise as $x$ approaches 
$x_{\mini}:=1/(a_{\mini}-a)<0$ from above.
Moreover, one can show that $g''(x)>0$ 
for all $x\in(x_{\mini},x_{\maxi})$,
i.e., $g(x)$ is a convex function.
Upon observing hat
$g(0)=1$, $g'(0)=a_{\mic}-a$
it thus follows that the equation
$g(x)=1$ is solved by $x=0$
for any given $a$ in (\ref{4}).
In the case $a=\amic$, there can be no further solution
of $g(x)=1$,
hence we set $y:=0$, and (\ref{6})-(\ref{8}) 
readily follow.
In the case $a\not=\amic$, there must exist exactly
one further $x\in(x_{\mini},x_{\maxi})\setminus\{0\}$
which solves $g(x)=1$, and this $x$-value
is now identified with $y$.
Again, (\ref{6}) and (\ref{7}) are easily verified,
while (\ref{8}) is recovered upon concluding from
(\ref{5}) that
$\sum_{n=1}^N [1+y\,(a-a_n)]p_n=1$.
With (\ref{7}) this yields
$1+y\,[a-\sum_{n=1}^N a_n\, p_n]=1$
and due to $y\not=0$ we arrive at (\ref{8}).

Note that the value of $y$ depends on 
the choice of $a$ in (\ref{4}).
It is thus sometimes more appropriate 
to view $y$ as a function $y(a)$ of $a$.
On the other hand, $a$ is often considered 
as arbitrary but fixed, hence it is justified 
to omit the $a$ dependence of $y$.

From (\ref{6}) and (\ref{7})
one can infer that
\begin{eqnarray}
\rho:=\sum_{n=1}^N p_n\ |n\rangle\langle n|
\label{9}
\end{eqnarray}
is a well defined density operator and that
\begin{eqnarray}
\rho^{1/2}:=\sum_{n=1}^N \sqrt{p_n}\ |n\rangle\langle n|
\label{10}
\end{eqnarray}
is Hermitian and satisfies $(\rho^{1/2})^2=\rho$.
Adopting the usual definition
\begin{equation}
f(A):=\sum_{n=1}^N f(a_n)\, |n\rangle\langle n| 
\label{11}
\end{equation}
for an arbitrary function $f(x):\RR\to\RR$, 
Eq. (\ref{9}) can be rewritten
by means of (\ref{5}) and (\ref{11}) as
\begin{eqnarray}
\rho=
\frac{1}{N}\,\frac{1}{1+y\,(a-A)} \ .
\label{12}
\end{eqnarray}

Next we consider an ensemble of uniformly 
distributed and normalized random vectors 
$|\psi\rangle\in\hr$.
In other words, the probability distribution
of those vectors $|\psi\rangle$ is invariant
under arbitrary unitary transformations
and thus all vectors $|\psi\rangle$ are 
equally likely.
For any given Hermitian operator 
$B:\hr\to\hr$ it follows
that the ``Hilbert space average'' (HA),
i.e., the average of $\langle\psi|B|\psi\rangle$
over the above ensemble of vectors 
$|\psi\rangle$, amounts to  \cite{bar09,gem04,llo88}
\begin{eqnarray}
\HA\left[\langle\psi|B|\psi\rangle\right]
=\frac{\tr\{B\}}{N} \ .
\label{13}
\end{eqnarray}
Likewise, the corresponding ``Hilbert space 
variance'' (HV) is given by \cite{bar09,gem04,llo88}
\begin{eqnarray}
& & \HV\left[\langle\psi|B|\psi\rangle\right]
:=
\HA\left[(\langle\psi|B|\psi\rangle
-
\HA\left[\langle\psi|B|\psi\rangle\right])^2\right]
\nonumber
\\
& & =
\frac{1}{N+1}\left(\frac{\tr\{B^2\}}{N}
-\left(\frac{\tr\{B\}}{N}\right)^2\right) \ .
\label{14}
\end{eqnarray}
The appearance of the denominator
$N+1$ on the right hand side of (\ref{14})
will play a key role in what follows.
A simple intuitive explanation of its 
origin is not know to this author, 
while the formal mathematical derivation
can be found, e.g., in 
Appendix C.1 of Ref. \cite{gem04}
or in the Appendix of Ref. \cite{llo88}.

Next we consider (not necessarily normalized)
vectors $|\varphi\rangle\in\hr$ of the form
\begin{eqnarray}
|\varphi\rangle := \sqrt{N} \rho^{1/2}|\psi\rangle
\label{15}
\end{eqnarray}
with $\rho^{1/2}$ from (\ref{10}).
For the ensemble of random vectors $|\psi\rangle$
as defined above, we thus obtain a
modified ensemble of vectors 
$|\varphi\rangle$ via (\ref{15}).
Denoting their ensemble average by an
overbar, we thus can conclude that
\begin{eqnarray}
\overline{\langle\varphi|B|\varphi\rangle}
=
\HA\left[
\langle\psi|N\,\rho^{1/2}B\rho^{1/2}|\psi\rangle
\right]
\label{16}
\end{eqnarray}
and with (\ref{13}) that
\begin{eqnarray}
\overline{\langle\varphi|B|\varphi\rangle}
=
\tr\{\rho^{1/2}B\,\rho^{1/2}\}
=
\tr\{\rho\, B\}
\ .
\label{17}
\end{eqnarray}
Likewise, the corresponding variance
\begin{eqnarray}
\sigma_B^2:=\overline{
\left(\langle\varphi|B|\varphi\rangle
-\overline{\langle\varphi|B|\varphi\rangle}\right)^2
}
\label{18}
\end{eqnarray}
can be rewritten with the help of (\ref{14}) as
\begin{eqnarray}
\sigma_B^2
& = &
\HV\left[
\langle\psi|N\,\rho^{1/2}B\rho^{1/2}|\psi\rangle
\right]
\nonumber
\\
& = &
\frac{
N\, \tr\{(\rho B)^2\}-\left(\tr\{\rho B\}\right)^2
}{N+1}
\ .
\label{19}
\end{eqnarray}
It readily follows that
\begin{eqnarray}
\sigma_B^2
\leq \tr\{(\rho B)^2\}
\ .
\label{20}
\end{eqnarray}
Considering $\tr\{C_1^\dagger C_2\}$
as a scalar product between two arbitrary
(not necessarily Hermitian) linear operators
$C_{1,2}:\hr\to\hr$, the Cauchy-Schwarz
inequality takes the form
$|\tr\{C_1^\dagger C_2\}|^2\leq
\tr\{C_1^\dagger C_1\}\tr\{C_2^\dagger C_2\}$.
Choosing $C_2=C_1^\dagger=\rho B$ it follows that
\begin{eqnarray}
\tr\{(\rho B)^2\}
\leq 
\tr\{(\rho B)(B\rho)\}
=
\tr\{\rho^2 B^2\}
\ .
\label{21}
\end{eqnarray}
Evaluating the trace by means of the 
eigenbasis of $B$, one finds
that 
\begin{eqnarray}
\tr\{\rho^2 B^2\}\leq \norm{B}^2\,  \tr\{\rho^2\}
\ ,
\label{22}
\end{eqnarray}
where $\norm{B}$ indicates the operator norm of
$B$ (largest eigenvalue in modulus).
Combining (\ref{20})-(\ref{22}) thus yields
\begin{eqnarray}
\sigma_B^2 & \leq & \norm{B}^2\, \QQ
\ ,
\label{23}
\\
\QQ & := & \tr\{\rho^2\}
= \sum_{n=1}^N p_n^2
\label{24}
\\
& = & \frac{1}{N^2}\sum_{n=1}^N\frac{1}{[1+y\,(a-a_n)]^2}
\label{25}
\\
& = & \frac{1}{N^2}\,\tr\left\{[1+y\,(a-A)]^{-2}\right\}
\label{26}
\end{eqnarray}
where we have exploited (\ref{5}), (\ref{9}),
and ({\ref{12}).

\subsection{Establishing dynamical typicality}
\label{s32}
%
Throughout this section we take for 
granted that $\QQ$ from (\ref{24}) 
satisfies
\begin{eqnarray}
\QQ\ll 1
\ .
\label{27}
\end{eqnarray}
A more detailed discussion of this 
assumption is postponed to the 
following sections.

Choosing for $B$ the identity operator, it follows from
(\ref{17})-(\ref{19}) and (\ref{24}) that
\begin{eqnarray}
& & \overline{\langle\varphi|\varphi\rangle} = 1
\ ,
\label{28}
\\
& & 
\overline{
\left(\langle\varphi|\varphi\rangle
-1\right)^2
}
=\frac{N \QQ-1}{N+1}
\leq \QQ \ .
\label{29}
\end{eqnarray}
Invoking the Chebyshev inequality from 
probability theory, one thus can infer that
\begin{eqnarray}
& & 
\pr\left(
| \langle\varphi|\varphi\rangle - 1 |\leq \QQ^{1/3}
\right)
\geq 1-\QQ^{1/3} \ ,
\label{29a}
\end{eqnarray}
where the left hand side denotes the probability
that $|\langle\varphi|\varphi\rangle -1 |\leq \QQ^{1/3}$
when randomly sampling vectors $|\varphi\rangle$ 
according to (\ref{15}).
Due to (\ref{27}), the overwhelming majority of all 
vectors $|\varphi\rangle$ in (\ref{15}) thus have 
norms very close to unity.

Choosing for $B$ the operator $A$ from
(\ref{3}), it follows with (\ref{8}),
(\ref{9}), and (\ref{17}) that
\begin{eqnarray}
& & \overline{\langle\varphi|A|\varphi\rangle} = a
\label{30}
\end{eqnarray}
and with (\ref{18}) and (\ref{23}) that
\begin{eqnarray}
\overline{
\left(\langle\varphi|A|\varphi\rangle
-a\right)^2
}\leq \norm{A}^2\, \QQ \ .
\label{31}
\end{eqnarray}
Clearly, adding an arbitrary constant $c\in\RR$ to $A$
and simultaneously to $a$ does not change the variance
on the left hand side of (\ref{31}), while $A$ on the
right hand side is replaced by $A+c\id$.
The best possible upper bound is thus
obtained by minimizing $\norm{A+c\id}$
over all $c\in\RR$. 
A straightforward calculation shows that
this minimal value will be given by $\da/2$,
where $\da$ is the measurement 
range of $A$ from (\ref{3a}).
Invoking Chebyshev's inequality once more, 
one thus can deduce from (\ref{31}) that
\begin{eqnarray}
\pr\left(
| \langle\varphi|A|\varphi\rangle - a |\leq \QQ^{1/3}\da/2
\right)
\geq 1-\QQ^{1/3} \ .
\label{31a}
\end{eqnarray}
In view of (\ref{27}), the overwhelming
majority of all vectors $|\varphi\rangle$ 
in (\ref{15}) thus exhibit expectation 
values $\langle\varphi|A|\varphi\rangle$, whose
deviations from 
the preset value $a$ in (\ref{4})
are very small compared 
to full range $\da$ over which those expectation 
values in principle could vary.

Finally, by choosing for $B$ the operator 
$O_t$ in (\ref{2}) and observing that 
$\norm{O_t}=\norm{O}$, one can infer
from (\ref{18}), (\ref{23}),
and (\ref{27})  along the very same 
line of reasoning as before that 
most
vectors 
$|\varphi\rangle$ in (\ref{15}) exhibit 
very similar expectation values 
$\langle \varphi|O_t|\varphi\rangle$.

So far, the initial states $|\varphi\rangle$ 
in (\ref{15}) are in general not 
normalized. But, as mentioned below
(\ref{29a}), the vast majority 
among them is almost of unit 
length. 
Hence, if we replace for every given
$|\psi\rangle$ the concomitant
$|\varphi\rangle$ in (\ref{15}) by 
its strictly normalized counterpart
\begin{eqnarray}
|\phi\rangle
:= 
\frac{|\varphi\rangle}{\sqrt{\langle\varphi|\varphi\rangle}}
=
\frac{\rho^{1/2}|\psi\rangle}
{\sqrt{\langle\psi|\rho |\psi\rangle}}
\ ,
\label{32}
\end{eqnarray}
then the ``new'' expectation values
$\langle\phi|A|\phi\rangle$
will mostly remain very close to the
``old'' ones, i.e., to
$\langle\varphi|A|\varphi\rangle$.
Essentially, this is a consequence of the
relation
\begin{eqnarray}
\langle\phi|A|\phi\rangle
=
\frac{\langle\varphi|A|\varphi\rangle}{\langle\varphi |\varphi\rangle}
\label{r1}
\end{eqnarray}
following from (\ref{32})
and of the fact that $\langle\varphi |\varphi\rangle$ is very close
to unity for most $|\varphi\rangle$'s according to (\ref{29a}).
More precisely, with the help of (\ref{r1}) and
\begin{eqnarray}
q(\varphi):=|1-\langle\varphi|\varphi\rangle|
\ ,
\label{b7a}
\end{eqnarray}
we can rewrite
$|\langle\phi|A|\phi\rangle - \langle\varphi|A|\varphi\rangle|$
as $q(\varphi) |\langle\phi|A|\phi\rangle|$.
Exploiting the triangle inequality, it follows that
\begin{eqnarray}
|\langle\phi|A|\phi\rangle -a| 
\leq 
q(\varphi)  |\langle\phi|A|\phi\rangle|
+
|\langle\varphi|A|\varphi\rangle -a| 
\ .
\label{b10}
\end{eqnarray}
Since $|\langle\phi|A|\phi\rangle|\leq \norm{A}$, 
an argument analogous to the one 
below (\ref{31}) then yields
\begin{eqnarray}
|\langle\phi|A|\phi\rangle -a| 
\leq 
q(\varphi) \da/2
+
|\langle\varphi|A|\varphi\rangle -a| 
\ .
\label{b11}
\end{eqnarray}
According to (\ref{31a}),
the probability that 
$|\langle\varphi|A|\varphi\rangle -a|\leq\QQ^{1/3}\da/2$
is at least $1-\QQ^{1/3}$,
and due to (\ref{29a}), (\ref{b7a}) 
the probability that $q(\varphi) \leq \QQ^{1/3}$
is at least $1-\QQ^{1/3}$.
Therefore, the probability that both
$|\langle\varphi|A|\varphi\rangle -a| \leq\QQ^{1/3}\da/2$
and 
$q(\varphi) \leq\QQ^{1/3}$
are simultaneously fulfilled
must be at least $1-2\QQ^{1/3}$.
Together with (\ref{b11}) we thus 
can conclude that
\begin{eqnarray}
\pr\left(
| \langle\phi|A|\phi\rangle - a |\leq \QQ^{1/3}\da
\right)
\geq 1-2\QQ^{1/3} \ .
\label{b12}
\end{eqnarray}

In conclusion, by sampling random 
initial conditions according to 
(\ref{32}), the vast majority of 
them exhibits expectation 
values  
$\langle\phi|A|\phi\rangle$
very close the preset value $a$
from (\ref{4}).
Along the same line of reasoning one
sees that for most of them also
the expectation values 
$\langle\phi|O_t|\phi\rangle$
in (\ref{1}) 
will be very similar.
Altogether, we thus recover 
dynamical typicality as announced 
below (\ref{2}).

As an aside, we note that the
possible values of 
$\langle\phi|A|\phi\rangle$
are bounded by the smallest and
largest eigenvalues of $A$,
hence the restriction of admissible 
$a$-values in (\ref{4}) still 
covers all physically meaningful 
cases.

\subsection{Necessary and sufficient conditions}
\label{s33}
According to (\ref{24}), $\QQ$ is the purity of 
$\rho$ and thus satisfies $0\leq \QQ\leq 1$.
The demonstration of dynamical typicality
in the preceding subsection is based on 
the additional assumption (\ref{27}).
In the opposite case that $\QQ$ is not small
(but still taking for granted $N\gg 1$, as required 
above Eq. (\ref{1})),
also the left hand side of (\ref{29})
is not any more a small quantity.
As a consequence, one expects that
the norm of most $|\varphi\rangle$'s is 
not any more close to unity
and the arguments below Eq. (\ref{32}) 
are no longer valid.
In particular, the denominator on the right 
hand side of (\ref{r1}) now exhibits non-negligible
random fluctuations, hence the same 
property will be inherited by the 
left hand side, since
there is no reason why the numerator
should (almost) exactly compensate 
the fluctuations of the denominator.
It is therefore reasonable to expect
that (\ref{27}) is in fact not only
a sufficient but also a necessary 
prerequisite for dynamical 
typicality,
provided the quite trivial cases 
with $\QQ$-values very close to 
unity are tacitly excluded.
Without providing the details we 
mention that this conjecture can 
also be confirmed more rigorously.

One readily concludes from 
(\ref{6}), (\ref{7}), and 
(\ref{24}) that 
\begin{eqnarray}
& & p^2_{\maxi} \leq  \QQ\leq p_{\maxi}
\ ,
\label{33}
\\
& & p_{\maxi}  :=  \max_n p_n \ .
\label{34}
\end{eqnarray}
The two relations in (\ref{33}) imply that $\QQ$ 
is small if and only if $p_{\maxi}$ is 
small.
As a consequence, (\ref{27}) is equivalent to
the following necessary and sufficient
condition for dynamical typicality:
\begin{eqnarray}
p_{\maxi}\ll 1 \ .
\label{35}
\end{eqnarray}

In passing we note that (\ref{7}) implies 
$p_{\maxi}\geq 1/N$, hence the condition
$N\gg 1$ (see above (\ref{1})) is automatically
guaranteed if (\ref{35}) is fulfilled.

\section{Comparison with Ref. \cite{mul11}}
\label{s4}
We recall that the ensemble of random
vectors $|\phi\rangle$ on the left hand 
side of (\ref{32}) derives from a uniformly 
distributed ensemble of 
normalized $|\psi\rangle$'s 
on the right hand side.
This specific ensemble of initial states
$|\phi\rangle$ was
demonstrated in Sec. \ref{s3} 
to exhibit dynamical typicality, 
provided condition (\ref{35}) is 
fulfilled.
In particular, the expectation value 
$\langle\phi|A|\phi\rangle$ 
turned out to be {\em almost} equal to the 
preset value $a$ from (\ref{4}) 
for {\em most} of those $|\phi\rangle$'s.
Apart from that, the actual properties
of the ensemble at hand are not very clear.
Specifically, it is not at all obvious 
how these findings for the 
ensemble 
(\ref{32})
can be translated into statements 
about the more natural but
quite different ensemble,
where {\em all} normalized 
vectors, whose expectation value is 
{\em strictly} equal to $a$, are 
realized with equal probability 
(and all other vectors are excluded).
It is exactly the latter ensemble
which is at the focus of the explorations 
by M\"uller, Gross, and Eisert in 
Ref. \cite{mul11}.
Accordingly,
our subsequent resolution of the
above issue will be largely based
on their results.

To begin with we note that 
the set of all possible vectors 
$|\psi\rangle$ on the right hand side 
of (\ref{15}) amounts to a unit 
sphere in $\CC^N$, or equivalently,
to a $2N-1$ sphere in $\RR^{2N}$.
Accordingly, the set of vectors
$|\varphi\rangle$ on the left hand side of
(\ref{15}) may be viewed as a $2N-1$
dimensional ellipsoid in $\RR^{2N}$.
After an additional normalization of 
each vector, see (\ref{32}),
the latter ellipsoid is deformed 
back into a $2N-1$ sphere,
henceforth denoted as
$S_{a}$, where the index $a$ 
refers to the preset $a$-value
from (\ref{4}). 

As mentioned above, the main focus 
in Ref. \cite{mul11} is put on a different 
set of 
vectors
$|\phi\rangle$, namely (in our present notation)
\begin{eqnarray}
M_{a}:=\{|\phi\rangle \ | \ \langle\phi|A|\phi\rangle=a \ 
\mbox{and}\ \langle\phi|\phi\rangle=1\} \ .
\label{36}
\end{eqnarray}
Due to the presence of {\em two} 
constraints on the right hand side,
the set $M_a$ generically amounts to 
a $2N-2$ dimensional submanifold of 
$\RR^{2N}$, namely the intersection of 
an ellipsoid and a sphere.
In other words, the two sets
$S_a$ and $M_a$ are quite 
different.

To go over from those two sets 
to the corresponding statistical
ensembles, we also need to specify 
the probability measures on each  
set.

The probability measure on $M_a$ considered
in Ref. \cite{mul11} is the
natural ``Hausdorff measure'' inherited 
from $\RR^{2N}$, i.e., the 
probability measure of any $2N-2$ 
dimensional  ``surface element'' on
$M_a$ is proportional to its ``size''.
In other words, the distribution may be
viewed as ``uniform'' on $M_a$:
Every vector $|\phi\rangle\in M_a$ is 
``equally likely'' in a very natural 
sense.

In contrast, the uniformly distributed
vectors $|\psi\rangle$ in (\ref{32}) generally
induce a quite non-trivial (far from uniform) probability 
measure on the above specified $2N-1$ dimensional
submanifold $S_a$.

From now on, the symbols $S_a$ and $M_a$
refer not only to the corresponding 
submanifolds (sets) but also to the above 
specified probability measures for each of 
them, i.e., they amount to full fledged 
statistical ensembles of normalized 
random vectors $|\phi\rangle$.

Given dynamical typicality as detailed in the
previous section is fulfilled for the ensemble
$S_a$,  does a similar property also hold for 
the ensemble $M_a$, and vice versa?
The answer to this highly non-trivial question
is ``yes'', as can be inferred from Section 6 
of Ref. \cite{mul11}.
However, the detailed arguments
and calculations are quite tedious 
and are therefore postponed to 
Appendix B.
In particular, it can be shown that
the main prerequisite in 
the previous section, namely the 
condition (\ref{35}),
is at the same time 
the crucial prerequisite in the
pertinent Theorem 1 of 
Ref. \cite{mul11}.
Moreover, the average of 
$\langle\phi|B|\phi\rangle$
over the ensemble $S_a$ agrees 
very well with the corresponding 
average over the ensemble 
$M_a$ for arbitrary $B$ and any 
given $a$-value, for which 
$\rho$ in (\ref{9}) satisfies 
condition (\ref{35}).

In conclusion, the two ensembles $S_a$ 
and $M_a$ are largely equivalent
as far as their dynamical typicality 
properties are concerned.
Since the ensemble $M_a$ is in some
sense considerably more ``natural''
than $S_a$, this equivalence notably
enhances the physical significance 
of the findings in the remainder 
of our paper.

The intuitive picture is that -- in spite
of the fact that the two sets $S_a$ and
$M_a$ are very different -- the extremely
non-uniform probability measure on $S_a$
essentially only leaves over vectors
$|\phi\rangle$ ``close to $M_a$''.

It should be emphasized that the
physical motivation and the 
interpretation of the results 
in Ref. \cite{mul11}
are very different from our present
dynamical typicality viewpoint.
Accordingly, our above conclusions are
not at all immediate consequences of 
the actual findings in Ref. \cite{mul11} 
itself, but rather should be considered 
as significant new insights of the 
present paper (see also Appendix B).

Technically speaking, the above mentioned
typicality result for the ensemble 
$M_a$ were established in Ref. \cite{mul11}
by means of concentration of 
measure concepts, and may be viewed
as an extension of Levy's Lemma 
\cite{pop06,pop06a}.
In particular, since $M_a$ in (\ref{36})
does not amount to a hypersphere, 
one cannot directly apply Levy's Lemma 
itself to this case.

\section{Exploring the conditions for dynamical typicality}
\label{s5}
%
\subsection{General considerations}
\label{s51}
In principle, once the observable 
$A$ in (\ref{3}) and the value of 
$a$ in (\ref{4}) 
have been fixed, 
$y$ is uniquely determined according 
to the discussion below
(\ref{8a}).
Hence, all $p_n$ in (\ref{5}) 
and their maximum in (\ref{34})
follow, and one can decide whether or not 
the necessary and sufficient condition for 
dynamical typicality in (\ref{35}) 
is fulfilled.
In practice, it is usually impossible 
to ``directly'' solve all the necessary 
equations, hence one has to 
figure out ``indirect criteria'' whether or 
not the given $A$ and $a$ satisfy (\ref{35}).
This is the main goal of the 
present section.

A first important special case arises 
by considering the microcanonical 
ensemble from (\ref{37}) and the 
concomitant microcanonical 
expectation value from (\ref{38}).
Namely, for the particular choice
$a=a_{\mic}$ we have
$y=0$  (see below (\ref{8a})),
and hence $\rho$ from (\ref{12}) 
must coincide with $\rhomic$.
In other words, we find that
\begin{eqnarray}
a=\amic
\Leftrightarrow
y=0
\Leftrightarrow
\rho=\rhomic
\Leftrightarrow
p_n=1/N\ \forall n\, .
\label{39}
\end{eqnarray}
Hence, (\ref{35}) is satisfied if an only 
if $N$ is large.
Moreover, (\ref{15}) and (\ref{32})
simplify to $|\phi\rangle=|\varphi\rangle=|\psi\rangle$, i.e., 
we recover the original ``non-dynamical''
or ``microcanonical'' typicality results 
from \cite{llo88,pop06,pop06a,gol06,sug07,sug12,tas16}
(see also \cite{rei07,gem04}).

From now on, we mainly focus on 
$a$-values with
\begin{eqnarray}
\amic < a < a_{\maxi} \ .
\label{40}
\end{eqnarray}
The corresponding results for 
$a_{\mini}<a < \amic$ are quite obvious 
(consider $-A$ instead of $A$)
and will only be briefly mentioned.

Denoting by $D_{\maxi}$ the multiplicity 
of the largest eigenvalue of $A$ 
(i.e., there are $D_{\maxi}$ indices 
$n$ with $a_n=a_{\maxi}$),
we can infer from (\ref{40}) 
and Eq. (\ref{a17}) in 
Appendix A that
\begin{eqnarray}
& & 0 < y \leq \frac{1-D_{\maxi}/N}{a_{\maxi}-a}  
\ . 
\label{41}
\end{eqnarray}
With the help of the first relation, $y>0$, in (\ref{41}),
one can conclude from (\ref{5}) and (\ref{34}) that
\begin{eqnarray}
& & p_{\maxi} = \frac{1}{N}\, \frac{1}{1-y\,(a_{\maxi}-a)} \ .
\label{42}
\end{eqnarray}

So far, $a$ in (\ref{4}) was considered 
as arbitrary but fixed. 
Next, we are interested in how things
change upon variation of $a$,
hence $y$ and thus $p_{\maxi}$ in
(\ref{42}) become functions of $a$
(see also the discussion 
above Eq. (\ref{9})).
A first general picture of how
$p_{\maxi}$ depends on $a$ within 
the range (\ref{40}) is provided 
by the following three observations:
First, $p_{\maxi}$ is an increasing 
function of $a$, as shown in Appendix C.
Second, $p_{\maxi}$ approaches $1/N$ 
for $a\to \amic$, see (\ref{39}).
Third, one readily confirms that 
when $a$ approaches $a_{\maxi}$, 
the unique $y$-value in 
(\ref{5}) compatible with
(\ref{6})-(\ref{8}) approaches 
$(1-D_{\maxi}/N)/(a_{\maxi}-a)$.
With (\ref{42}) it follows that
\begin{eqnarray}
p_{\maxi}\to 1/D_{\maxi}\ \ 
\mbox{for $a\to a_{\maxi}$.}
\label{43}
\end{eqnarray}

If $D_{\maxi}\gg 1$ we can 
conclude \cite{f1}
that the condition (\ref{35}) 
is satisfied for {\em any} $a$ within 
the range (\ref{40}).
(If $a_{\mini}<a<\amic$ the same conclusion 
holds on condition that the smallest 
eigenvalue of $A$ is highly degenerate.)

From now on, we restrict ourselves 
to cases, for which $D_{\maxi}\gg 1$ is 
{\em not} satisfied, including the 
most common case $D_{\maxi}=1$.
In view of (\ref{43}), $a$-values 
sufficiently close to $a_{\maxi}$ thus 
inevitably must lead to a violation of
(\ref{35}).
``How closely'' $a$ may approach $a_{\maxi}$
without violating (\ref{35}) depends on 
many details of the spectrum of $A$,
as we will see in the following.

\subsection{Eigenvalue density approximation}
\label{s52}
To begin with, we assume that the spectrum 
of $A$ gives rise to an approximately 
constant density of eigenvalues 
close to its upper limit $a_{\maxi}$.
More precisely, when rewriting 
(\ref{5}), (\ref{7}) as
\begin{eqnarray}
& & 
\int dx \ w(x)\, 
\frac{1}{1+y\,(a-x)}=1
\ ,
\label{44}
\\
& & 
w(x) := \frac{1}{N} \sum_{n=1}^N \delta(a_n-x)
\label{45}
\end{eqnarray}
we assume that 
$w(x)$ 
can be approximated by some constant, 
positive value $w_0$ for $x$ close 
to $a_{\maxi}$.
Moreover, we assume that $a$ is sufficiently 
close to $a_{\maxi}$ so that the latter
approximation applies to all $x\in[a,a_{\maxi}]$.
Note that the function $w(x)$ is normalized 
to unity and thus may be viewed as an
eigenvalue probability distribution.
A more rigorous approach, which does not
exploit the above assumptions, is worked
out in Appendix D, see also
Sec. \ref{s54} below.

As a first consequence of the 
above assumptions,
the number of eigenvalues of $A$
which are larger than (or equal to)
$a$ is given by
\begin{eqnarray}
N_a:=N \, \int_{a}^{a_{\maxi}}dx\ w(x)
\label{46}
\end{eqnarray}
and can be approximated as
\begin{eqnarray}
N_a \simeq N\, (a_{\maxi}-a)\, w_0 \ .
\label{47}
\end{eqnarray}
Second, upon
restricting the integration domain in (\ref{44})
to $x\geq a$ and observing that $w(x)=0$ for $x>a_{\maxi}$
we can conclude that 
\begin{eqnarray}
\int_{a}^{a_{\maxi}}dx \ w(x)\, 
\frac{1}{1+y\,(a-x)}
\leq 1
\ .
\label{48}
\end{eqnarray}
Approximating $w(x)$ by $w_0$ and
performing the integration yields
\begin{eqnarray}
\frac{w_0}{y}\,
\ln\left(\frac{1}{1-y\,(a_{\maxi}-a)}\right)
\leq 1 \ .
\label{49}
\end{eqnarray}
With (\ref{42}) it follows that
\begin{eqnarray}
\frac{w_0}{y}\,\ln(p_{\maxi}N) \leq 1
\label{50}
\end{eqnarray}
and with (\ref{47}) that
\begin{eqnarray}
\frac{N_a}{N}\,\frac{\ln(p_{\maxi}N)}{y\,(a_{\maxi}-a)} 
\leq 1
\ .
\label{51}
\end{eqnarray}
By taking into account (\ref{41}), the 
inequality (\ref{51}) implies
\begin{eqnarray}
\ln(p_{\maxi}N) \leq \frac{N}{N_a}\ .
\label{52}
\end{eqnarray}
Observing that $\alpha,\,\beta\in\RR$ satisfy 
$\alpha\leq\beta$ if and only if $e^\alpha\leq e^\beta$
it follows that
\begin{eqnarray}
p_{\maxi}\leq \frac{1}{N}\exp\left\{\frac{N}{N_a}\right\}
\ .
\label{53}
\end{eqnarray}
Under the assumption that
\begin{eqnarray}
N_a\gg \frac{N}{\ln N}
\label{54}
\end{eqnarray}
one readily concludes that $(1/N)\exp\{N/N_a\}\ll 1$
and with (\ref{53}) that
$p_{\maxi}\ll 1$.

Altogether we find that (\ref{35})
is guaranteed under the following sufficient
conditions: The number $N_a$ of eigenvalues
$a_n$ larger than $a$ must satisfy (\ref{54})
and the density of levels between $a$ and
$a_{\maxi}$ must be (approximately) constant.
For large $N$ one expects that in many
cases there will be $a$-values close
to $a_{\maxi}$ which satisfy both 
conditions.
Recalling that $p_{\maxi}$ is an increasing 
function of $a$ (see Appendix C), 
it follows that
if the above conditions are met by some
$a$ close to $a_{\maxi}$, then (\ref{35}) 
remains valid also for all smaller 
$a$-values down to $\amic$.
A similar conclusion applies for $a<\amic$
provided the eigenvalues $a_n$ of $A$
are sufficiently numerous and of 
approximately constant density 
near $a_{\mini}$.

Referring to Appendix D for the detailed
calculations,
we remark that when $a$ approaches $a_{\maxi}$
so closely that (\ref{54}) is no longer
satisfied then one can show that 
$p_{\maxi}$ is no longer small.
In other words, if the 
eigenvalue probability distribution
$w(x)$ assumes a constant, positive value
near $a_{\maxi}$, then (\ref{54})
(together with an analogous 
property near $a_{\mini}$)
is not only a sufficient but also a necessary
condition for (\ref{35}).

\begin{figure}
\epsfxsize=1.0\columnwidth
\epsfbox{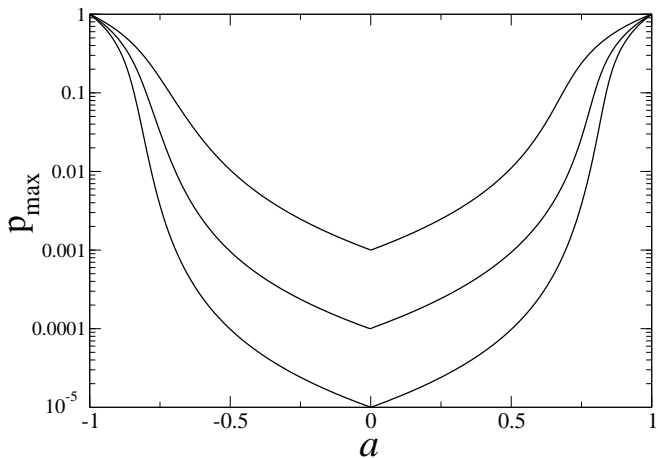}
\caption{\label{fig1}
Dependence of $p_{\maxi}$ from (\ref{42}), (\ref{55}) 
on the parameter $a$ in (\ref{4}), (\ref{5}) by 
numerically solving (\ref{7}) for $y$
(see also Sec. \ref{s6})
with $N=10^2$, $N=10^3$, and
$N=10^4$ (top down).
In each case, the $N$ eigenvalues $a_n$ in (\ref{5})
were randomly generated so that the difference 
between neighboring eigenvalues are Wigner 
distributed \cite{bro81}
and $a_{\mini}=-1$, $a_{\maxi}=1$.
}
\end{figure}

A numerical illustration of these analytical
predictions is provided in Fig. 1. 
It exemplifies the case when $w(x)$
can be well approximated by $w_0=1/2$ 
for all $x\in(a_{\mini},a_{\maxi})=(-1,1)$
in the sense explained below (\ref{45}).
Note that $\amic$ in (\ref{38})
is always very close to zero in
these examples.
For $a$-values close to $a_{\maxi}=1$,
we see that $p_{\maxi}$ in 
Fig. \ref{fig1} becomes small 
only for very large $N$
(and likewise for $a$ close to $a_{\mini}=-1$).
To better understand this numerical finding,
we temporarily consider the inequalities 
in (\ref{48})-(\ref{53}) as approximate
equalities.
With (\ref{47}) we thus can rewrite (\ref{53})
as $p_{\maxi}\simeq N^{-1}\exp\{2\,(1-a)^{-1}\}$.
For a given (fixed) value of $p_{\maxi}$ 
(e.g. $p_{\maxi}=0.01$) the 
corresponding value of 
$1-a$ thus scales as $1/\ln N$, i.e., 
$a$ approaches $a_{\maxi}=1$
only very slowly upon 
increasing $N$.
Closer inspection shows that
(\ref{48})-(\ref{53}) indeed become
approximate equalities in the 
above considered case that
both $p_{\maxi}$ and $1-a$ are small.
We may also recall that it is the purity
$P$ which actually counts in the 
dynamical typicality considerations
in Sec. \ref{s32}, and that $P$ is
bounded by $p_{\maxi}$ according
to (\ref{33}). Quantitatively, $P$ may
thus be considerably smaller than 
$p_{\maxi}$. However, also $P$ will
still decrease very slowly with $N$
(at most as $(1/\ln N)^2$).

In passing we note that  $p_\maxi$ is given 
by (\ref{42}) for $a\geq\amic$, while for 
$a\leq\amic$ the corresponding formula
reads
\begin{eqnarray}
p_{\maxi} = \frac{1}{N}\frac{1}{1-y\,(a-a_{\mini})}
\ .
\label{55}
\end{eqnarray}
As a consequence, $p_\maxi$
turns out to be a continuous but not differentiable 
function of $a$ at $a=\amic$,
see Fig. 1.

\subsection{Large $N$ limit}
\label{s53}
So far we considered the Hilbert 
space dimension $N$ as 
large but fixed.
In order to draw conclusions about
how things (in particular $p_\maxi$) change 
upon variations of $N$, it is necessary (and sufficient)
to specify how the spectrum of the
observable $A$ in (\ref{3}) changes 
with $N$,
which {\em a priori} appears to be a
quite subtle problem in itself.
A physically natural way would be to model
how the ``same'' measurement device
acts on different systems.
Here, we adopt an alternative approach
which is physically somewhat abstract 
but mathematically quite common:
Namely, we are mainly interested in
the large $N$ limit and we assume that 
the eigenvalue probability distribution
from (\ref{45}) approaches a well-defined
limit for asymptotically large $N$
(with slightly washed-out delta functions
on the right hand side of (\ref{45})).
In particular, $a_{\maxi}$ and $a_{\mini}$ are
assumed to become (asymptotically) $N$-independent.
Moreover, the symbol $w(x)$ is henceforth
used for this limiting function, and
likewise for $w_0$ in (\ref{47}) and
(\ref{50}).
For an example, see also Fig. 1.

For any given $a$ close to $a_{\maxi}$, 
the solution $y$ of (\ref{44})
is thus (asymptotically) $N$-independent.
From (\ref{50}) it follows that 
$p_{\maxi}$ approaches zero as $N$ 
tends to infinity.
Since $p_{\maxi}$ is monotonically increasing with $a$
(see above or Appendix C),
we thus find that (\ref{35}) will be 
satisfied for any given $a$-value 
from (\ref{40}) when $N\to\infty$.

Yet another conclusion is mentioned 
here without detailed proof since 
it is quite plausible in view of 
the so far results:
If the eigenvalue probability distribution 
$w(x)$ tends to zero
for $x\to a_{\maxi}$ then (\ref{35}) is 
violated for all $a$ beyond some threshold 
$a_{\mathrm{th}}<a_{\maxi}$. 
Analogous conclusions apply to the 
lower end of the spectrum of $A$.
Closely related rigorous results are
derived in Appendix D.

A particularly prominent example arises 
when the spectrum satisfies a so-called 
semicircle law, e.g. due to a 
corresponding random matrix 
character of $A$ \cite{bro81}.
It seems reasonable to expect that after
the projection (restriction) of the original 
observable to an energy shell as described at the
end of Sec. \ref{s2}, the resulting reduced
observable $A$ may possibly exhibit 
such features reminiscent of a random matrix.

Further instructive examples are obtained 
by considering any model of 
$M$ identical subsystems with negligible 
interactions and finite dimensional 
Hilbert spaces.
Namely, when choosing any observable of 
the form $A=(1/M)\sum_{m=1}^M A_m$,
where the $A_m$ act on the single 
subsystems but are otherwise identical,
one can infer that
$w(x)$ approaches $\delta(\amic-x)$ 
for $M\to\infty$ by exploiting
the central limit theorem.
It follows that (\ref{35}) is 
violated for {\em any} $a\not=\amic$.
A particular example of this type
is originally due to Ref. \cite{mul11} 
(see Sec. 3 therein).
However, it should be emphasized 
that the energy of the system is 
not confined to some non-trivial 
energy window in these examples,
in contrast to the physical 
setup we mainly have in mind 
in the present paper
(see end of Sec. \ref{s2}).

\subsection{Outlook}
\label{s54}
Generally speaking, the idea to replace
the exact distribution in (\ref{45}) 
by a ``washed out'' approximation 
appears quite reasonable and also 
the so obtained results seem decent.
However, in cases where the levels $a_n$
are very far from being approximately
equally spaced (at least locally), 
such a smooth local eigenvalue probability 
distribution may still be well-defined, 
but the conclusions may become questionable.
An extreme example arises when 
the levels $a_n$ can be partitioned
into groups, so that each group contains
the same number $D$ of elements and
the eigenvalues $a_n$ are equal within 
each group ($D$-fold degeneracies).
If the levels belonging to different 
groups are still approximately equally 
spaced, then a smoothened distribution $w(x)$
still appears to be reasonably well-defined 
($D$ possibly large but fixed, 
$N$ sufficiently large so that the gaps are 
still small).
However, if $D\gg 1$ we know 
from the discussion below (\ref{43}) 
that (\ref{35}) will be satisfied 
without any further restriction regarding $a$ 
and $a_n$, in contradiction to the
above conclusions under the 
assumption of a smooth $w(x)$.

In view of such concerns, some further
results are derived in Appendix D
by means of more rigorous,
but technically also more involved
calculations.

In summary, the range of $a$-values, 
for which dynamical typicality applies,
crucially depends on the detailed
spectral properties of $A$,
in particular near the upper and 
lower ends of the spectrum.

\section{Further remarks and comparison with Ref. \cite{bar09}}
\label{s6}
Conceptually, a dynamical typicality result
is particularly satisfying when referring to
the set of normalized vectors from (\ref{36}), i.e., 
when it amounts to a statement 
about all initial states
$|\phi\rangle$ whose expectation values 
$\langle \phi|A|\phi\rangle$ are 
exactly equal to $a$. 

As detailed in Ref. \cite{mul11} and
in Sec. \ref{s3}, the set of vectors
$|\phi\rangle$ arising via (\ref{32}) is
very different from the set (\ref{36}), 
and also the probability measures 
on the two sets are very different.
Nevertheless, the typicality statements
for the two ensembles of random vectors 
are largely equivalent.

In turn, the main virtue of (\ref{32}) 
is its constructive character: 
It is an explicit recipe to generate 
``typical initial states'' $|\phi\rangle$.
The only remaining disadvantage is
that all eigenvalues and eigenvectors
of $A$ are explicitly needed, 
see (\ref{5}) and (\ref{10}).
Moreover, in order to determine 
$y$, the transcendental 
equation $g(x)=1$ has to be solved 
as discussed below (\ref{8a}).
Since $g(x)$ in (\ref{8a}) consists
of $N$ summands, the numerical 
effort to solve $g(x)=1$ scales 
linearly with $N$.
Generically, the diagonalization of 
$A$ is thus numerically much more 
expensive than the determination 
of $y$.
At least for $a$-values sufficiently
close to $a_{\mic}$, this diagonalization
of $A$ as well as the numerical 
determination of $y$
can be circumvented as follows.

Considering $y=y(a)$ as
a function of $a$, our first goal
is to expand $y(a)$ about $a_{\mic}$.
From (\ref{39}), we can infer 
that $y(a_{\mic})=0$.
Focusing on small $\delta a:=a-a_{\mic}$,
it follows that $y(a)$ is small 
and (\ref{5})  can be rewritten as 
\begin{eqnarray}
p_n=\frac{1}{N}\sum_{k=0}^\infty [-y(a)\, (a-a_n)]^k \ .
\label{56}
\end{eqnarray}
Replacing $a$ by $a_{\mic}+\delta a$ and
introducing the expansion of $y(a)$ about
$a_{\mic}$ on the right hand side of (\ref{56}),
plugging in the resulting expressions
for the $p_n$ into (\ref{7}), and finally 
comparing terms with equal powers 
of $\delta a$, one obtains equations
for the derivatives of $y(a)$ at $a=a_{\mic}$
which can be iteratively solved.
The resulting series expansion 
of $y(a)$ reads
\begin{eqnarray}
y(a) & = & 
\frac{1}{m_2}(a-a_{\mic})-\frac{m_3}{m_2^3}(a-a_{\mic})^2 + ...
\ ,
\label{57}
\\
m_k & := & 
\frac{1}{N}\sum_{n=1}^N(a_n-a_{\mic})^k
=\frac{1}{N}\tr\{(A-a_{\mic})^k\}
\ ,
\qquad
\label{58}
\end{eqnarray}
where we exploited (\ref{3}) and (\ref{11})
in the last identity.
In the same vein, one can deduce from (\ref{5}) 
for small values of $y=y(a)$ the expansion
\begin{eqnarray}
\sqrt{p_n}=
\frac{
1 -\frac{1}{2}y(a)\, (a-a_n)+\frac{3}{8}[y(a)\, (a-a_n)]^2+...
}{N^{1/2}}
\ .
\label{59}
\end{eqnarray}
Accordingly, (\ref{10}) takes the form
\begin{eqnarray}
\rho^{1/2}=\frac{
1 +\frac{1}{2}y(a)\, (A-a)+\frac{3}{8}[y(a)\,(A-a)]^2+...
}{N^{1/2}}
\ .
\label{60}
\end{eqnarray}
For $a$-values sufficiently close to $a_{\mic}$, 
one can introduce (\ref{57}) into (\ref{60}) and
truncate the series after a few terms, yielding very 
good approximations for the exact ensemble 
in (\ref{32}) without the need to diagonalize 
the operator $A$.

The latter expansions amounts to a systematic
generalization of the original dynamical typicality
explorations by Bartsch and Gemmer in Ref. 
\cite{bar09}.
Indeed, as they pointed out below
their formula (3), their approach is
meant to be restricted to small 
values of their so-called disequilibrium 
parameter $d$, 
which in turn is essentially equivalent 
to our present expansion parameter
$\delta a$
\cite{f2}.
Accordingly, beyond the linear approximation
of (\ref{57}) and (\ref{60}) adopted 
by Bartsch and Gemmer \cite{bar09}, 
notable differences may generically
arise between their ensemble
of initial states $|\phi\rangle$
and either of the two ensembles 
from (\ref{32}) and (\ref{36}),
as considered in our present work
and 
in Ref. \cite{mul11}, respectively.

\section{Summary}
\label{s7}
By unifying and extending previous works 
(mainly Refs. \cite{bar09,mul11}, but also 
Refs. \cite{fin09,rei07,alv08}),
dynamical typicality has been established in the sense that
the vast majority of all
pure states, which initially exhibit the same expectation 
value $a$ of some given observable $A$, also exhibit 
very similar expectation values of the same or some 
different observable $O$ after evolving according to 
the Schr\"odinger equation up to some later time point $t$.
More precisely, our main focus was on the preconditions
under which such a dynamical typicality property holds true.
In general, we found that those conditions depend in a very
complicated way on the spectrum of $A$ and on the 
value of $a$, but not on $O$ or $t$, nor
on the Hamiltonian $H$ which governs
the dynamics.
In particular, the range of $a$-values, for which dynamical
typicality applies, depends crucially on the spectral
properties of $A$ near the largest and smallest 
eigenvalues, denoted as $a_{\maxi}$ and $a_{\mini}$,
respectively.
For instance, if the eigenvalues 
$a_{\maxi}$ and $a_{\mini}$ are both 
highly degenerate, then the entire range 
$[a_{\mini},a_{\maxi}]$ of physically 
reasonable $a$-values is admitted.
As the density of eigenvalues near 
$a_{\maxi}$ and $a_{\mini}$ decreases, 
larger and larger neighborhoods of 
$a_{\maxi}$ and $a_{\mini}$ no longer
belong to the interval of admitted $a$-values.
However, for sufficiently large
dimensions $N$ of the considered Hilbert space,
those excluded vicinities of $a_{\maxi}$ and 
$a_{\mini}$ may still become arbitrarily
small in many cases.
In other cases, for instance if the
spectrum of $A$ obeys a semi-circle law,
those excluded vicinities of
$a_{\maxi}$ and $a_{\mini}$ may remain 
finite even for asymptotically large 
dimensions $N$.
In any case, for sufficiently large (but finite)
$N$, there always exists at least some small
interval of admitted $a$-values around 
the microcanonical expectation value of
$A$, or equivalently, the average over all 
eigenvalues of $A$.

\begin{acknowledgments}
Numerous extremely enlightening discussions
with Jochen Gemmer and Ben N. Balz are
gratefully acknowledged.
Special thanks is due to Lennart Dabelow
for carefully reading the manuscript.
This work was supported by the 
Deutsche Forschungsgemeinschaft (DFG)
under Grant No. RE 1344/10-1 and
within the Research Unit FOR 2692
under Grant No. RE 1344/12-1.
\end{acknowledgments}

\appendix
\section{}
%
As in the main text, we consider an 
observable $A$ of the form (\ref{3}),
and we denote by $\amic$
its microcanonical 
expectation value from (\ref{38}),
by $a_{\maxi}$ and $a_{\mini}$ its
largest and smallest 
eigenvalues, and by $D_{\maxi}$ 
and $D_{\mini}$ the degeneracy
of $a_{\maxi}$ and $a_{\mini}$,
respectively
(i.e., there are $D_{\maxi}$ indices
$n$ with $a_n=a_{\maxi}$, 
and likewise for $D_{\mini}$).
Introducing the open interval
\begin{equation}
I_A:=(a_{\mini},a_{\maxi}) \ ,
\label{a2}
\end{equation}
we choose an arbitrary but 
fixed $a\in I_A$ and define 
the interval
\begin{eqnarray}
I_a:=\left(\frac{-1}{a-a_{\mini}},\, \frac{1}{a_{\maxi}-a}\right)
\ .
\label{a3}
\end{eqnarray}
From now on, the symbol $x$ indicates a real 
variable and we always silently take for 
granted that its value is 
restricted according to
\begin{eqnarray}
x\in I_a \ .
\label{a4}
\end{eqnarray}

It follows that the
\begin{eqnarray}
q_n(x) & := & \frac{1}{1+x(a-a_n)}
\label{a5}
\end{eqnarray}
are well defined, positive functions
for all $x\in I_a$, and likewise for
the function $g(x)$ defined in 
(\ref{8a}).
From these definitions, 
one readily can infer that
\begin{eqnarray}
g(0) & = & 1
\ ,
\label{a7}
\\
g'(0) & = & \amic -a
\ ,
\label{a8}
\\
g(x) & \to & \infty \ \ \mbox{for}\ 
x \uparrow \frac{1}{a_{\maxi}-a}
\ ,
\label{a9}
\\
g(x) & > & 1 \ \ \mbox{for}\ 
x> \frac{1-D_{\maxi}/N}{a_{\maxi}-a}
\ ,
\label{a10}
\\
g(x) & \to & \infty \ \ \mbox{for}\ 
x\downarrow -\frac{1}{a-a_{\mini}}
\ ,
\label{a11}
\\
g(x) & > & 1 \ \ \mbox{for}\ 
x < -\frac{1-D_{\mini}/N}{a-a_{\mini}}
\ ,
\label{a12}
\\
g''(x) & = & \frac{2}{N}\sum_{n=1}^N (a-a_n)^2 q^3_n(x) >0
\ \mbox{for all}\ 
x\in I_a , \ \qquad\ 
\label{a13}
\end{eqnarray}
where $x\in I_a$ is tacitly assumed in (\ref{a10}) and (\ref{a12}).

Next we consider the equation
\begin{equation}
g(x)=1 
\label{a14}
\end{equation}
for an arbitrary but fixed $a\in I_A$.
According to (\ref{a7}), one solution 
is $x=0$.
If $a\not=\amic$, then (\ref{a7})-(\ref{a13}) 
imply that there exists exactly one further solution 
of (\ref{a14}) within the interval
$I_a$,
henceforth denoted as $y(a)$.
Moreover, this solution $y(a)$ satisfies
\begin{eqnarray}
-\frac{1-D_{\mini}/N}{a-a_{\mini}}\leq y(a)\leq \frac{1-D_{\maxi}/N}{a_{\maxi}-a}
\label{a15}
\end{eqnarray}
and the sign of $y(a)$ must agree with the
sign of $a-\amic$.
In particular, if $a=\amic$ then $y(a):=0$ 
is the only solution of (\ref{a14}).

In summary, for any given $a\in I_A$ there exists
a unique real number $y(a)\in I_a$ with the properties
\begin{eqnarray}
\!\!\!\!\!\!\!\!\!\!\!\!
& & a=\amic \ \Leftrightarrow \ y(a) = 0
\ ,
\label{a16}
\\
\!\!\!\!\!\!\!\!\!\!\!\!
& & 
a \in (\amic,\,a_{\maxi})\ \Rightarrow
\ 0< y(a) \leq\frac{1-D_{\maxi}/N}{a_{\maxi}-a}
\ ,
\label{a17}
\\
\!\!\!\!\!\!\!\!\!\!\!\!
& & 
a \in (a_{\mini},\, \amic)\ \Rightarrow
\ 0>y(a)>-\frac{1-D_{\mini}/N}{a -a_{\mini}}
\ ,
\label{a18}
\\
\!\!\!\!\!\!\!\!\!\!\!\!
& & p_n(a) := 
\frac{1}{N}\frac{1}{1+y(a)\,(a-a_n)}>0\ \mbox{for all} \ n 
\ ,
\qquad
\label{a19}
\\
\!\!\!\!\!\!\!\!\!\!\!\!
& & \sum_{n=1}^N p_n(a) = 1
\ .
\label{a20}
\end{eqnarray}
It follows that
\begin{eqnarray}
1
& = & 
\sum_{n=1}^N [1+y(a)(a-a_n)]\, p_n(a)
\nonumber
\\
& = & 
[1+y(a)a]\sum_{n=1}^N p_n(a)
-y(a)\sum_{n=1}^N a_n p_n(a)
\nonumber
\\
& = & 
1+y(a)\left[a-\sum_{n=1}^N a_n p_n(a)\right]
\ .
\label{a21}
\end{eqnarray}
If $y(a)\not=0$, we can conclude that
\begin{eqnarray}
\sum_{n=1}^N a_n p_n(a) = a \ .
\label{a22}
\end{eqnarray}
If $y(a)=0$, the same result follows 
from (\ref{38}) and (\ref{a16}).

In turn, one readily sees that there exists no
further real number $y(a)$ so that all three 
properties (\ref{a19}), (\ref{a20}), and (\ref{a22})
are still satisfied. Omitting the argument
$a$ in $y(a)$ and in $p_n(a)$, one thus recovers 
the statement from Eqs. (\ref{4})-(\ref{8}) 
in the main text.

\section{}
This appendix provides the omitted details 
in the comparison at the end of 
Sec. \ref{s4} between our present 
approach and the one from Ref. \cite{mul11}.

We start out from the submanifolds 
$S_a$ and $M_a$ and the probability measure 
on each of them, see Sec. \ref{s4}.
In our present work, we demonstrate that
dynamical typicality, as specified in 
Sec. \ref{s3}, holds for $S_a$ if 
condition (\ref{35}) is satisfied.
In Ref. \cite{mul11} it is shown that the 
same kind of typicality holds for $M_a$ 
on conditions which 
are specified in Theorem 1 therein.
Moreover, one can infer from Section 6 of 
Ref. \cite{mul11} that dynamical typicality 
for $M_a$ implies the same for $S_a$.
In particular, the ensemble averaged 
expectation values of any given observable 
$B$ are very close to each other for the 
two ensembles $M_a$ and $S_a$.
Strictly speaking, three additional steps are
required to arrive at this conclusion:
(i) The notation in Ref. \cite{mul11}
has to be translated into our present
notation. This will be done in the second
succeeding paragraph.
(ii) Rather than the ensemble $S_a$, which 
is generated via (\ref{32}), the ensemble
generated via (\ref{15}) is actually
considered in Ref. \cite{mul11}.
Their equivalence has been pointed
out below Eq. (\ref{32}).
(iii) One has to show that the mapping 
of a vector $|\varphi\rangle\in\hr$ to its
expectation value $\langle\varphi|B|\varphi\rangle$
amounts to a Lipschitz continuous function
(see p. 820 in \cite{mul11}).
This has been demonstrated, e.g., 
in Ref. \cite{pop06a}
(see Lemma 5 therein).

In the remainder of this appendix
we show that the prerequisites
and/or the applicability of Theorem 1 in Ref. \cite{mul11} 
are essentially tantamount to the condition (\ref{35}) 
required by our present approach.
The details of the pertinent Theorem 1 in Ref. \cite{mul11}
are quite involved and therefore not reproduced 
here. Most of the subsequent discussion can be 
followed without knowledge of those details.

To begin with, we point out the main notational differences:
Our present $A=\sum_{n=1}^N a_n\, |n\rangle\langle n|$
corresponds to $H=\sum_{k=1}^n E_k\, |k\rangle\langle k|$
in Ref. \cite{mul11}. Likewise, our 
$a_{\maxi}$, $a_{\mini}$, and $a$ correspond to 
$E_{\maxi}$, $E_{\mini}$, and $E$, respectively.
Furthermore, $M_E$ and $E_A$ in Ref. \cite{mul11}
correspond to $M_a$ from (\ref{36}) and $\amic$
from (\ref{38}), respectively.
Similarly to Theorem 1 in \cite{mul11},
only $a<\amic$ are considered 
from now on (the case $a\geq \amic$
then readily follows, see 
also below Eq. (\ref{40})).
Finally, the definitions $E':=E+s$ and $E_k':=E_k+s$
from \cite{mul11} are translated into $a+s$ and
$a_n+s$, respectively.

According to Equation (2) in Ref. \cite{mul11},
the parameter $s$ is defined as the solution of 
the equation
\begin{eqnarray}
E' & = & (1+\delta_E)\, \left[\frac{1}{n}\sum_{k=1}^n\left(E'_k\right)^{-1}\right]^{-1}
\ ,
\label{b1}
\\
\delta_E & := &
\frac{1}{n}+\frac{\epsilon}{\sqrt{n}}\left(1+\frac{1}{n}\right)
\ ,
\label{b2}
\end{eqnarray}
where $\epsilon$ will be specified later.
Furthermore, $s$ must satisfy $s>-E_{\mini}$.
Introducing
\begin{eqnarray}
y(a,\delta):=-\frac{1}{a+s} \ ,
\label{b3}
\end{eqnarray}
these conditions translate into
\begin{eqnarray}
& & 
g(y(a,\delta)) = (1+\delta)
\ ,
\label{b4}
\\
& & 
\delta :=
\frac{1}{N}+\frac{\epsilon}{\sqrt{N}}\left(1+\frac{1}{N}\right)
\ ,
\label{b5}
\\
& & 
0>y(a,\delta)>-\frac{1}{a-a_{\mini}}
\label{b6}
\ ,
\end{eqnarray}
where $g(x)$ is defined in (\ref{8a}).
For $\delta\to 0$ we thus recover (\ref{a14})
and hence $y(a,\delta)$ goes over into $y(a)$
from (\ref{a15})-(\ref{a20}).

The most important condition in Theorem 1 of 
Ref. \cite{mul11} reads $\epsilon>E'/E_Q'$
with $E'_Q:=\left(\frac{1}{n}
\sum_{k=1}^n [E'_k]^{-2}\right)^{-1/2}$ 
and translates along 
the same lines as above into
\begin{eqnarray}
\frac{1}{N}\sum_{n=1}^N 
\left[q_n(y(a,\delta))\right]^2
< \epsilon^2 
\label{b7}
\end{eqnarray}
with $q_n(x)$ from (\ref{a5}).
This condition suggests that the so far 
unspecified parameter $\epsilon$ should 
be chosen as large as possible.
Since $\epsilon$ also enters 
via (\ref{b4}), (\ref{b5}), a somewhat
more elaborate consideration is needed 
to show that the latter statement
remains true nevertheless.
Finally, the definition $c:=3E'_{\mini}/(32 E')$
in \cite{mul11} translates into 
$c=3/[32 N p_{\maxi}(a,\delta)]$, where
$p_{\maxi}(a,\delta):=\max_n q_n(y(a,\delta))/N$.
Since $y(a,\delta)<0$, see (\ref{b6}),
it follows that
$p_{\maxi}(a,\delta)=N^{-1}[1+y(a,\delta)(a-a_{\mini})]^{-1}$.
The largest possible choice of $\epsilon$
so that the exponential term in Equation
(1) of Ref. \cite{mul11} is 
-- as requested -- a large positive number
even for quite small $t$-values
turns out to be $\epsilon=\tilde\epsilon/
[\sqrt{N}p_{\maxi}(a,\delta)]$,
where $\tilde\epsilon$ is $N$ 
independent and satisfies 
$\tilde\epsilon\ll 1$.
For large $N$ it follows that $\delta$ 
in (\ref{b5}) can be approximated as zero,
hence $y(a,\delta)\simeq y(a)$ (see above)
and $p_{\maxi}(a,\delta) \simeq p_{\maxi}(a)$ 
(see (\ref{55}) and (\ref{c7})).
Accordingly, by exploiting 
(\ref{12}), (\ref{24}), 
(\ref{a5}), and (\ref{a19})
the condition (\ref{b7}) can be rewritten as
\begin{eqnarray}
N^2 p_{\maxi}^2 \QQ \ll 1 \ .
\label{b8}
\end{eqnarray}
With (\ref{33}) it follows that
\begin{eqnarray}
p_{\maxi}\ll 1/N^{2/3} 
\label{b9}
\end{eqnarray}
will be a somewhat stronger condition
(i.e., a sufficient but possibly not
necessary condition for (\ref{b8})
to hold true).
For large $N$, this is a slightly
(but not dramatically much) stronger
condition than the requirement (\ref{35})
of our present approach.
In turn, when $p_{\maxi}$ is not small
then it immediately follows with (\ref{33})
that (\ref{b8}) is {\em not} satisfied.

Comparing (\ref{35}) and (\ref{b9}) shows
that the present approach is actually 
valid under slightly more general conditions than
the corresponding Theorem 1 in Ref. \cite{mul11},
but for practical purposes this generalization 
seems only of minor relevance.

\section{}
In this appendix, we show that $p_{\maxi}$ is a
monotonically increasing function of $a$
for $a>\amic$, and monotonically decreasing 
for $a<\amic$.

By differentiating (\ref{a20}) with respect to
$a$ and introducing (\ref{a19}) one can infer that
\begin{eqnarray}
\sum_{n=1}^N
\frac{1}{N}\frac{y'(a)(a-a_n)+y(a)}{[1+y(a)\,(a-a_n)]^2}
= 0
\ .
\label{c1}
\end{eqnarray}
Multiplying this equation by $y(a)$ and rewriting
the resulting numerator on the left hand side
as $y'(a)[1+y(a)(a-a_n)-1]+y^2(a)$ yields
\begin{eqnarray}
& & \sum_{n=1}^N
\frac{1}{N}\frac{y'(a)}{1+y(a)\,(a-a_n)}
=[y'(a)-y^2(a)]S(a)
\ , \qquad
\label{c2}
\\
& & 
S(a):=\sum_{n=1}^N
\frac{1}{N}\frac{1}{[1+y(a)\,(a-a_n)]^2}
\ .
\label{c3}
\end{eqnarray}
With (\ref{a19}) and (\ref{a20}) we
can rewrite (\ref{c2}) as
\begin{eqnarray}
& & y'(a)[S(a)-1] = y^2(a)S(a) \ .
\label{c4}
\end{eqnarray}
From (\ref{a19}) and (\ref{c3})
we can conclude that
\begin{eqnarray}
S(a)=N\sum_{n=1}^N p_n^2(a)=N\,\QQ \ ,
\label{c5}
\end{eqnarray}
where $\QQ$ is the purity of $\rho$ from 
Eqs. (\ref{24}) of the main text.
Observing (\ref{a19}) and (\ref{a20}) one 
can infer that $S(a)=1$ if $y(a)=0$ 
and $S(a)>1$ in any other case.
(Equivalently, the purity $\QQ$ takes its 
minimal possible value $1/N$ if an only 
if $\rho$ is the microcanonical 
density operator.)
With (\ref{a16}) we thus can conclude that
\begin{eqnarray}
y'(a) = y^2(a)\frac{S(a)}{S(a)-1} >0\ \mbox{for all $a\not=\amic$.}
\label{c6}
\end{eqnarray}
For continuity reason it follows that
$y'(\amic)\geq 0$
(see also (\ref{57})).

Next we address the dependence of the quantity
\begin{eqnarray}
p_{\maxi}(a):=\max_n p_n(a) 
\label{c7}
\end{eqnarray}
on the argument $a\in I_A$.
We first focus on $a$-values so that 
(\ref{a17}) is fulfilled.
Exploiting (\ref{a19}) it follows that
\begin{eqnarray}
p_{\maxi}(a) = \frac{1}{N}\frac{1}{1-y(a)(a_{\maxi}-a)}
\ .
\label{c8}
\end{eqnarray}
Differentiating and introducing (\ref{c6}) 
yields
\begin{eqnarray}
p'_{\maxi}(a) & = &  \frac{1}{N}
\frac{y'(a)(a_{\maxi}-a)-y(a)}{[1-y(a)(a_{\maxi}-a)]^2}
\nonumber
\\
& = & 
\frac{N\, p^2_{\maxi}(a)\, y(a)\, h(a)}{S(a)-1}
\label{c9}
\ ,
\\
h(a) & := & y(a)\,S(a)(a_{\maxi}-a)-(S(a)-1)
\nonumber
\\
& = & 1-S(a)[1-y(a)(a_{\maxi}-a)] \ .
\label{c10}
\end{eqnarray}
With (\ref{c3}) we can conclude that
\begin{eqnarray}
& & S(a)[1-y(a)(a_{\maxi}-a)]
=
\sum_{n=1}^N 
p_n(a)\,b_n(a)
\ ,
\qquad
\label{c11}
\\
& & b_n(a) := \frac{1-y(a)(a_{\maxi}-a)}{1-y(a)(a_n-a)}
\nonumber
\\
& & =\frac{p_{\maxi}(a)}{p_{\maxi}(a)+y(a)(a_{\maxi}-a_n)/N}
\ ,
\label{c12}
\end{eqnarray}
where we exploited (\ref{c8}) in the last step.
Observing that $y(a)>0$, $p_{\maxi}(a)>0$, 
and $a_{\maxi}-a_n\geq 0$ for all $n$
it follows that $0\leq b_n(a)\leq 1$
and with (\ref{a20}) that
$\sum_{n=1}^N p_n(a)\, b_n(a)\in [0,1]$.
With (\ref{c10}) and (\ref{c11}) we can conclude that 
$h(a)\geq 0$ and with (\ref{c9}) that
\begin{eqnarray}
p'_{\maxi}(a)\geq 0 \ \mbox{for all $a\in (\amic,a_{\maxi})$}\ ,
\label{c13}
\end{eqnarray}
i.e., $p_{\maxi}(a)$ is a monotonically 
increasing function of $a$.
For $a$-values so that (\ref{a18}) is fulfilled,
one finds in the same way that
\begin{eqnarray}
p'_{\maxi}(a)\leq 0 \ \mbox{for all $a\in (a_{\mini},\amic)$}\ ,
\label{c14}
\end{eqnarray}
i.e., $p_{\maxi}(a)$ is a monotonically decreasing
function of $a$.

\section{}
In this Appendix, the necessary and
sufficient condition (\ref{35})
for dynamical typicality is explored
by means of more rigorous calculations
than those presented in 
Sec. \ref{s52}.

Without loss of generality we assume 
that the eigenvalues $a_n$ of $A$
from (\ref{3}) are ordered by 
magnitude:
\begin{eqnarray}
a_1\leq a_2\leq \ldots \leq a_N=:a_{\maxi}\ .
\label{d10}
\end{eqnarray}
As in Sec. \ref{s51}, we can and will focus on 
$a$-values within the range from (\ref{40}),
implying (\ref{41}) and (\ref{42}). 

Similarly as below Eq. (\ref{43}), we furthermore 
disregard the trivial case that $D_{\maxi}\gg 1$, 
i.e., we take for granted that the largest 
eigenvalue of $A$ is {\em not} highly degenerate.
As a consequence, the case $a\to a_{\maxi}$ is then also 
trivial: (\ref{35}) is always violated according 
to (\ref{43}).
We thus can and will restrict ourselves 
to the case
\begin{eqnarray}
\Delta a:= a_{\maxi}-a>0 \ .
\label{d30}
\end{eqnarray}

From (\ref{41}) and (\ref{d30}) it follows that
\begin{eqnarray}
\delta:= y\,\Delta a>0 \ .
\label{d31}
\end{eqnarray}
Moreover, (\ref{42}) can be rewritten as
\begin{eqnarray}
p_{\maxi}=\frac{1}{N}\frac{1}{1-\delta} \ .
\label{d32}
\end{eqnarray}

Finally, we restrict ourselves
to cases where $a$ coincides with 
one of the eigenvalues $a_n$, say
\begin{eqnarray}
a=a_{\nu}
\ .
\label{d20}
\end{eqnarray}
Since $p_{\maxi}$ is an increasing function 
of $a$ (see Appendix C), this is a quite minor
restriction: If we can show that (\ref{35})
is fulfilled for $a=a_\nu$, then the
same conclusion remains true for all
$a\leq a_\nu$ (as long as (\ref{40})
is fulfilled).
Likewise, if (\ref{35}) is violated
for $a=a_\nu$, then the same 
applies for all $a\geq a_\nu$.

\subsection*{D1. Derivation of a sufficient condition}
Our first goal will be to derive a sufficient
criterion under which (\ref{35}) is fulfilled. 
To begin with, we define the linear function
\begin{eqnarray}
f(x):=a_{\maxi}-b\, (N-x) \ ,
\label{d40}
\end{eqnarray}
where the slope $b$ is chosen as follows
\begin{eqnarray}
b & := & \max_{\nu\leq n<N}b_n
\ ,
\label{d41}
\\
b_n & := & \frac{a_N-a_n}{N-n}
\ .
\label{d42}
\end{eqnarray}
One thus can conclude that
\begin{eqnarray}
f(n)\leq a_n\ \mbox{ for all $n\in\{\nu,...,N\}$}
\label{d50}
\end{eqnarray}
and that (\ref{d41}) is the smallest possible
slope in (\ref{d40}) which exhibits this
property.
In other words, $f(n)$ represents a
tight linear lower bound for all eigenvalues
$a_n\geq a$ with the extra constraint 
that $f(N)=a_{\maxi}$.

Note that the slope of a linear function
through the two points $a=a_\nu$ and
$a_{\maxi}=a_N$ would be given by $b_\nu$.
The ratio between the two slopes,
\begin{eqnarray}
\beta := b/b_\nu
\label{d70}
\end{eqnarray}
quantifies how well the true $a_n$'s 
(with $\nu\leq n\leq N$) can be
approximated by a linear function in the above
sense of a rigorous lower bound.
In particular, one readily sees that
$\beta\geq 1$.
On the other hand, not too large 
$\beta$'s may be expected in 
many cases.

In the following, we will 
show that
\begin{eqnarray}
N_a\gg \beta\, \frac{N}{\ln N}
\label{d90}
\end{eqnarray}
is a sufficient condition for  
(\ref{35}),
where $\beta$ is defined in
(\ref{d70}) and where,
analogously as in (\ref{46}),
\begin{eqnarray}
N_a:=N-\nu
\label{d80}
\end{eqnarray}
is the number of eigenvalues $a_n$
which are larger than $a$.
This is the rigorous counterpart 
of the heuristic condition (\ref{54}) 
in the main text.

In order to prove this claim, we exploit 
(\ref{5})-(\ref{7}) to conclude
\begin{eqnarray}
1\geq S:=\sum_{n=\nu}^N p_n 
=
\frac{1}{N}\sum_{n=\nu}^N\frac{1}{1+y\,(a-a_n)} \ .
\label{d100}
\end{eqnarray}
Defining an auxiliary function $a(x)$ via
\begin{eqnarray}
a(x):=a_n\ \mbox{for $x\in(n-1,n]$, $n=1,...,N$}\ .
\label{d105}
\end{eqnarray}
it follows that
\begin{eqnarray}
S=\frac{1}{N}\int_{\nu-1}^N dx\, \frac{1}{1+y\,(a-a(x))} \ .
\label{d109}
\end{eqnarray}
Since all summands on the right hand side of 
(\ref{d100}) are positive (see (\ref{6})), 
the same applies to the integrand on the right hand 
side of (\ref{d109}),
implying that
\begin{eqnarray}
S \geq \frac{1}{N}\int_{x_0}^Ndx\, \frac{1}{1+y\,(a-a(x))} 
\label{d110}
\end{eqnarray}
for any
\begin{eqnarray}
x_0\in[\nu-1,N] \ .
\label{d111}
\end{eqnarray}
From (\ref{d50}) and (\ref{d105}) 
one can deduce that $f(x)\leq a(x)$
for all $x\in(\nu-1,N]$.
Since $y> 0$ (see (\ref{41}))
it follows that 
$1+y\,(a-f(x))\geq 1+y\,(a-a(x))$.
Hence we can conclude from (\ref{d110})  that
\begin{eqnarray}
S \geq  \frac{1}{N}\int_{x_0}^Ndx\, \frac{1}{1+y\,(a-f(x))} \ .
\label{d120}
\end{eqnarray}
Recalling that $f(x)$ in (\ref{d40}) 
is a linear function,
the integral on the right hand side can be readily 
evaluated.
Choosing $x_0$ so that $f(x_0)=a$
one can infer from (\ref{d40})
that $x_0$ is uniquely fixed in this way and
from (\ref{d20}), (\ref{d50}) that (\ref{d111}) 
is satisfied.
By means of a straightforward  but somewhat 
lengthy calculation and exploiting the
relations (\ref{d30}), (\ref{d31}), 
(\ref{d40}), (\ref{d42}), (\ref{d70}), (\ref{d80})
one finally obtains
\begin{eqnarray}
S\geq \frac{N_a}{N\beta \delta} \,
\ln\left(\frac{1}{1-\delta}\right) \ .
\label{d130}
\end{eqnarray}

It remains to be shown that if (\ref{d90})
is taken for granted, then (\ref{35}) 
follows:
Given (\ref{d90}) is valid, 
and observing that $N_a<N$ and 
$\beta \geq 1$, it follows that $N\gg 1$. 
If $\delta$ is not close to unity,
say $\delta\leq 1-1/\sqrt{N}$,
then (\ref{d32}) implies that 
$p_{\maxi}\leq 1/\sqrt{N}$, i.e.,
(\ref{35}) is fulfilled. 
Hence we are left with the case $\delta>1-1/\sqrt{N}$.
Exploiting (\ref{d32}) in the argument of the logarithm
in (\ref{d130}) and approximating the remaining
factor $\delta$ by unity, we can conclude with 
(\ref{d100}) and (\ref{d130}) that
\begin{eqnarray}
1\geq \frac{N_a}{N\beta} \,
\ln\left(p_{\maxi}N\right) =:Q \ .
\label{d140}
\end{eqnarray}
Solving for $p_{\maxi}$ yields
\begin{eqnarray}
p_{\maxi}=\frac{1}{N}\exp\left\{Q N \beta/N_a
\right\}
\ .
\label{d150}
\end{eqnarray}
It follow that $p_{\maxi}\ll 1$, i.e. (\ref{35}), is
fulfilled if and only if 
$\exp\left\{Q N \beta/N_a\right\}\ll N$.
Taking logarithms on both sides, and exploiting that
$N\gg 1$, this is tantamount to
$Q N \beta/N_a\ll \ln N$ and hence to
$Q N \beta/\ln N\ll N_a$.
Since $Q\leq 1$ according to (\ref{d140}),
the latter relation is guaranteed by our 
premise (\ref{d90}).

\subsection*{D2. Lower bound for $p_{\maxi}$}
We consider the following subsets of $G:=\{1,..,N\}$:
\begin{eqnarray}
G_1 & := & \{ n \in G\, | \, a_n \leq  a-\Delta a\}
\ ,
\label{d01}
\\
G_2 & := & \{ n \in G\, |\, a_n >  a-\Delta a \ 
\mbox{and}\ a_n \leq a \}
\ ,
\label{d02}
\\
G_3 & := & \{ n \in G\, |\, a_n > a\}
\ .
\label{d03}
\end{eqnarray}
Hence, $G$ is the disjoint union 
of the subsets $G_\kappa$, $\kappa=1,...,3$.
Denoting the number of elements
contained in the subset $G_\kappa$
as
\begin{eqnarray}
N_\kappa := |G_\kappa| \ ,
\label{d03a}
\end{eqnarray}
it follows that
\begin{eqnarray}
N_1+N_2+N_3=N
\ .
\label{d04}
\end{eqnarray}
In particular, $N_3$ is identical to $N_a$
from (\ref{46}) and (\ref{d80}),
\begin{eqnarray}
N_3=N-\nu=N_a \ .
\label{d81}
\end{eqnarray}
Moreover, defining
\begin{eqnarray}
S_\kappa := \sum_{n\in G_\kappa}p_n
\label{d82}
\end{eqnarray}
we can conclude with (\ref{7}) that
\begin{eqnarray}
S_1+S_2+S_3 =1 
\ .
\label{d83}
\end{eqnarray}
Finally, (\ref{5}), (\ref{d30}), 
(\ref{d31}), and (\ref{d01})-(\ref{d03}) 
can be exploited to conclude
\begin{eqnarray}
p_n & \leq & \frac{1}{N}\frac{1}{1+\delta}\ \mbox{for all}\ n\in G_1
\ ,
\label{d06}
\\
p_n & \leq & \frac{1}{N} \ \mbox{for all}\ n\in G_2
\ ,
\label{d07}
\\
p_n & \leq & \frac{1}{N}\frac{1}{1-\delta}\ \mbox{for all}\ n\in G_3
\ .
\label{d08}
\end{eqnarray}
With (\ref{d03a}) and (\ref{d82}) it follows that
\begin{eqnarray}
S_1 & \leq & \frac{N_1}{N}\frac{1}{1+\delta}
\ ,
\label{d160}
\\
S_2 & \leq & \frac{N_2}{N}
\ ,
\label{d170}
\\
S_3 & \leq & \frac{N_3}{N}\frac{1}{1+\delta}
\ .
\label{d180}
\end{eqnarray}
In combination with (\ref{d04}) 
and (\ref{d83}) this yields after a short
calculation the result
\begin{eqnarray}
S_1+S_2+S_3 & = & 1 \leq 1 + \frac{\delta}{N(1-\delta^2)}\ R
\ ,
\label{d190}
\\
R & := & \delta(N_1+N_3)-N_1+N_3  
\ .
\label{d200}
\end{eqnarray}
With (\ref{d31}) and (\ref{d32}) it follows
that $R$ in (\ref{d190}) must be non-negative 
and with (\ref{d200}) that
\begin{eqnarray}
\delta\geq\frac{N_1-N_3}{N_1+N_3}
\ .
\label{d210}
\end{eqnarray}

For instance, if $a$ is sufficiently close 
to $a_{\maxi}$ so that $N_3\leq N_1/3$ then
(\ref{d210}) implies $\delta \geq 1/2$
and with (\ref{d32}) we obtain the 
lower bound $p_{\maxi}\geq 2/N$.
Upon further increasing $a$ towards $a_{\maxi}$,
the ratio $N_3/N_1$ decreases,
and the lower bounds for $\delta$ and
$p_{\maxi}$ increase towards limits
not too far from their largest
possible values compatible 
with (\ref{41}) and (\ref{43}).

\subsection*{D3. Derivation of a necessary condition}
The lower bound for $p_{\maxi}$ implied 
by (\ref{d210}) implicitly
amounts to a necessary criterion for 
(\ref{35}), which, however does not
substantially go beyond the findings 
around (\ref{43}) in the main text.
The basic idea in the following considerations
is to proceed like in the previous Sec. D2,
except that a better upper bound for $S_3$
than in (\ref{d180}) will be derived.
The latter in turn is achieved by 
similar arguments as in Sec. D1.

To begin with, we define the function
\begin{eqnarray}
f(x) := a_{\maxi} - b\, (N-1-x)^\gamma
\label{d300}
\end{eqnarray}
with an arbitrary but fixed exponent
\begin{eqnarray}
\gamma >0 \ ,
\label{d310}
\end{eqnarray}
and where the prefactor $b$ is chosen as follows
\begin{eqnarray}
b & := & \min_{\nu\leq n <N} b_n
\ ,
\label{d320}
\\
b_n & := & \frac{a_N-a_n}{(N-n)^\gamma}
\ .
\label{d330}
\end{eqnarray}
It follows that
\begin{eqnarray}
f(n)\geq a_{n+1}\ \mbox{for all $n\in\{\nu-1,..,N-1\}$}
\label{d340}
\end{eqnarray}
and hence
\begin{eqnarray}
f(x)\geq a(x) \ \mbox{for all $x\in [\nu-1,\,N-1]$}
 \ ,
\label{d350}
\end{eqnarray}
where $a(x)$ is defined in (\ref{d105}).
Similarly as in (\ref{d70}) we define
\begin{eqnarray}
\sigma :=b/b_\nu
\label{d360}
\end{eqnarray}
but now it follows from (\ref{d320}) 
and (\ref{d330}) that $\sigma \leq 1$.
For the rest, the intuitive meaning of 
$f(x)$ and $\sigma $ is analogous to the
discussion below (\ref{d50}).

With (\ref{5}) and (\ref{34})
it follows that $p_N=p_{\maxi}$ 
and with (\ref{d10}), (\ref{d20}),
(\ref{d03}), and (\ref{d82})
that
\begin{eqnarray}
S_3 & = & p_{\maxi}+S_3'
\ ,
\label{d370}
\\
S_3' & := & \frac{1}{N}\sum_{n=\nu+1}^{N-1}\frac{1}{1+y\,(a-a_n)}
\ .
\label{d380}
\end{eqnarray}
Similarly as in (\ref{d109})-(\ref{d120}) 
one can exploit (\ref{d350}) to conclude
\begin{eqnarray}
S_3' \leq \frac{1}{N}\int_{x_0}^{N-1}dx \, \frac{1}{1+y\,(a-f(x))}
\label{d390}
\end{eqnarray}
for any $x_0\leq\nu$. In particular, 
we can choose $x_0$ so that $f(x_0)=a$.
Replacing the integration variable $x$ 
via substitution by $u:=(N-1-x)/\eta$, where
\begin{eqnarray}
\eta:=
N_3 \left(\frac{1-\delta}{\sigma  \delta}\right)^{1/\gamma}
 \ ,
\label{d400}
\end{eqnarray}
it follows with 
(\ref{d30}), (\ref{d31}), (\ref{d81}),
(\ref{d300}), (\ref{d330}), (\ref{d360})
that
\begin{eqnarray}
S_3' & \leq & \frac{\eta}{N(1-\delta)}\,
 I\left(\left[\frac{\delta}{1-\delta}\right]^{1/\gamma}\right)
\ ,
\label{d410}
\\
I(v) & := & \int_0^v du\,\frac{1}{1+u^\gamma} \ .
\label{d420}
\end{eqnarray}

For $\gamma=1$, i.e., the function $f(x)$ from 
(\ref{d300}) amounts to a linear upper bound 
in (\ref{d340}),
one finds with (\ref{d400})-(\ref{d420})
that
\begin{eqnarray}
S_3'\leq \frac{N_3}{N\sigma \delta}\ln\left(\frac{1}{1-\delta}\right)
\ .
\label{d440}
\end{eqnarray}
Introducing (\ref{d04}), (\ref{d160}), (\ref{d170}), (\ref{d370}),
(\ref{d400}) into (\ref{d83}) implies
\begin{eqnarray}
\!\!\!\!\!\!\!\!\!\!
& & 
\frac{N_1+N_2+N_3}{N}=S_1+S_2+S_3
\nonumber
\\
\!\!\!\!\!\!\!\!\!\!
& & 
\leq
\frac{N_1}{N}\frac{1}{1+\delta}+\frac{N_2}{N}+p_{\maxi}+
\frac{N_3}{N\sigma \delta}\ln\left(\frac{1}{1-\delta}\right)
\, .
\label{d450}
\end{eqnarray}
By means of (\ref{d32}) we thus obtain
\begin{eqnarray}
p_{\maxi} \geq 
\frac{N_1}{N}\frac{\delta}{1+\delta}+
\frac{N_3}{N}\left(1-\frac{\ln(p_{\maxi}N)}{\sigma \delta}\right)
 \ .
\label{d460}
\end{eqnarray}

In the following, we restrict ourselves 
to the case 
\begin{eqnarray}
N_2,\, N_3\leq N_1/3 \ .
\label{d470}
\end{eqnarray}
With (\ref{d04}) and (\ref{d210}) it follows that
\begin{eqnarray}
N_1 & \geq & 3\, N/5
\ ,
\label{d480}
\\
\delta &\geq & 1/2 \ .
\label{d490}
\end{eqnarray}
One readily verifies that this implies $\delta/(1-\delta)\geq 1/3$ 
and with (\ref{d460}) and $p_{\maxi}\leq 1$ (see (\ref{7}))
that
\begin{eqnarray}
p_{\maxi} &  \geq &
\frac{1}{5} - 
\frac{N_3}{N}\frac{2\ln N }{\sigma }
=(1-R)/5
\ ,
\label{d500}
\\
R & := & \frac{N_3}{N}\frac{10}{\sigma }\ln N
\ .
\label{d510}
\end{eqnarray}
It follows that a necessary prerequisite for (\ref{35}) is
$R\geq1/2$ and thus
\begin{eqnarray}
N_3\geq \frac{\sigma }{2}\frac{N}{\ln N}
\ .
\label{d520}
\end{eqnarray}

Upon comparison of the sufficient condition (\ref{d90}) for
(\ref{35}) with the necessary condition (\ref{d520})
(and recalling (\ref{d81})), 
we see that not much room is left in between
the two conditions, provided the $a_n$'s near $a_{\maxi}$
allow for a reasonable linear approximation,
i.e., $\beta$ in (\ref{d70}) is not much larger than
unity and $\sigma $ in (\ref{d360}) (with $\gamma=1$)
not much smaller than unity.
In other words, the breakdown of (\ref{35}) 
is expected to happen when
$N_a=N_3$ is roughly comparable to $N/\ln N$.
The extra condition for $N_3$ in (\ref{d470}) is 
then automatically fulfilled, and usually the same is 
expected to apply for $N_2$ in (\ref{d470}).
In other words, our above derivation of (\ref{d520}) 
is self consistent in the non-trivial regime, i.e.,
whenever the stronger condition (\ref{d90}) is violated.

Finally, we turn to exponents $\gamma$ 
in (\ref{d300}) with $1>\gamma>0$ 
(see also (\ref{d310}); 
exponents $\gamma>1$ are of minor interest).
As before, we focus on the case (\ref{d480}).
Hence, $\delta$ satisfies (\ref{d490}) 
and the argument of $I$ in (\ref{d410})
is lower bounded by unity.
Observing that in (\ref{d420}) 
the integrand is upper 
bounded by unity for $u\leq 1$ and by 
$u^{-\gamma}$ for $u\geq 1$
then yields
\begin{eqnarray}
I(v)
\leq 
1+\frac{v^{1-\gamma}-1}{1-\gamma}
=
\frac{v^{1-\gamma}-\gamma}{1-\gamma}
\leq
\frac{v^{1-\gamma}}{1-\gamma}
\ .
\label{d530}
\end{eqnarray}
Introducing (\ref{d400}) and (\ref{d530}) into
(\ref{d410}) results in
\begin{eqnarray}
S_3'
& \leq &
\frac{N_3}{N(1-\gamma) \sigma^{1/\gamma} \delta}
\leq
\frac{N_3}{N}\frac{2}{(1-\gamma)\sigma^{1/\gamma}} \ ,
\label{d540}
\end{eqnarray}
where we exploited (\ref{d490}) in the last step.
Similarly as in (\ref{d450})-(\ref{d500}) one
can conclude that
\begin{eqnarray}
p_{\maxi} &  \geq & (1-R')/5
\ ,
\label{d550}
\\
R' & := & \frac{N_3}{N}\frac{10}{(1-\gamma)\sigma^{1/\gamma}}
\ .
\label{d560}
\end{eqnarray}
It follows that a necessary prerequisite for (\ref{35}) is
$R'\geq1/2$ and thus
\begin{eqnarray}
N_3\geq \frac{(1-\gamma)\sigma^{1/\gamma}}{2}\, N
\ .
\label{d570}
\end{eqnarray}
In contrast to the result (\ref{d520}) for $\gamma=1$,
the result (\ref{d570}) for $\gamma<1$ predicts a 
violation of (\ref{35}) at a certain fraction $N_3/N$ 
of eigenvalues $a_n$ near $a_{\maxi}$, which
does not approach zero 
even in the limit $N\to\infty$.



\begin{thebibliography}{11}
\bibitem{f1}
A simpler way to arrive at the same 
conclusion is as follows:
Summing on the left hand side of 
(\ref{7}) only over the $D_{\maxi}$ 
indices $n$ with $a_n=a_{\maxi}$
yields $D_{\maxi}\, p_{\maxi}\leq 1$.
Thus $D_{\maxi}\gg 1$ implies
$p_{\maxi}\ll 1$.
\bibitem{f2}
With respect to the $m_k$
from (\ref{58}), it is assumed in 
Ref. \cite{bar09} that $a_{\mic}=0$ and
$m_2=1$, which can always be
achieved by means of a trivial
``renormalization'' of $A$.
In addition, it is required in \cite{bar09}
that the $m_k$ are at most of the order 
of unity (compared to $N$)
for all $k=3,...,8$.
This is a non-trivial restriction
with respect to $A$ and notably
stronger than our present conditions
in Sec. \ref{s5} and Appendix D.
\end{thebibliography}
\end{document}